\documentclass[aps,prl,reprint,showpacs,twocolumn,superscriptaddress]{revtex4}

\usepackage[dvips]{graphicx}
\usepackage{dcolumn}
\usepackage{bm}
\usepackage{xspace}
\usepackage{multirow}
\usepackage{natbib}
\usepackage{color}

\begin{document}

\preprint{}
\title{Topological semimetal in a Bi-Bi$_2$Se$_3$ infinitely adaptive superlattice phase}

\author{T. Valla}
\email{valla@bnl.gov}
\affiliation{Condensed Matter Physics and Materials Science Department, Brookhaven National Lab, Upton, New York 11973, USA}
\author{Huiwen Ji}
\affiliation{Department of Chemistry, Princeton University, Princeton, New Jersey 08544, USA}
\author{L. M. Schoop}
\affiliation{Department of Chemistry, Princeton University, Princeton, New Jersey 08544, USA}
\author{A. P. Weber}
\affiliation{National Synchrotron Light Source, Brookhaven National Lab, Upton, New York 11973}
\affiliation{Department of Physics, University of Missouri-Kansas City, Kansas City, Missouri 64110, USA}
\author{Z.-H. Pan}
\affiliation{Condensed Matter Physics and Materials Science Department, Brookhaven National Lab, Upton, New York 11973, USA}
\author{J. T. Sadowski}
\affiliation{Center for Functional Nanomaterials, Brookhaven National Laboratory, Upton, New York 11973, USA}
\author{E. Vescovo}
\affiliation{National Synchrotron Light Source, Brookhaven National Lab, Upton, New York 11973}
\author{A. V. Fedorov}
\affiliation{Advanced Light Source, Lawrence Berkeley National Laboratory, Berkeley, California 94720, USA}
\author{A. N. Caruso}
\affiliation{Department of Physics, University of Missouri-Kansas City, Kansas City, Missouri 64110, USA}
\author{Q. D. Gibson}
\affiliation{Department of Chemistry, Princeton University, Princeton, New Jersey 08544, USA}
\author{L. M\"uchler}
\affiliation{Institut f\"ur Anorganische und Analytische Chemie, Johannes Gutenberg-Universitat, 55128 Mainz, Germany}
\author{C. Felser}
\affiliation{Institut f\"ur Anorganische und Analytische Chemie, Johannes Gutenberg-Universitat, 55128 Mainz, Germany}
\affiliation {Max-Planck-Institut f\"ur Chemische Physik fester Stoffe, 01187 Dresden, Germany.}
\author{R. J. Cava}
\email{rcava@princeton.edu}
\affiliation{Department of Chemistry, Princeton University, Princeton, New Jersey 08544, USA}
\date{\today}

\begin{abstract}
We report spin- and angle-resolved photoemission studies of a topological semimetal from the infinitely adaptive series between elemental Bi and Bi$_2$Se$_3$. The compound, based on Bi$_4$Se$_3$, is a 1:1 natural superlattice of alternating Bi$_2$ layers and Bi$_2$Se$_3$ layers; the inclusion of S allows the growth of large crystals, with the formula Bi$_4$Se$_{2.6}$S$_{0.4}$. The crystals cleave along the interfaces between the Bi$_2$ and Bi$_2$Se$_3$ layers, with the surfaces obtained having alternating Bi or Se termination. The resulting terraces, observed by photoemission electron microscopy, create avenues suitable for the study of one-dimensional topological physics. The electronic structure, determined by spin- and angle- resolved photoemission spectroscopy, shows the existence of a surface state that forms a large, hexagonally shaped Fermi surface around the $\Gamma$ point of the surface Brillouin zone, with the spin structure indicating that this material is a topological semimetal. 

\end{abstract}
\vspace{1.0cm}


\maketitle
\pagebreak
Three-dimensional (3D) topological insulators (TIs) have Dirac-like surface states in which the spin of the electron is locked perpendicular to its momentum in a chiral spin structure where electrons with opposite momenta have opposite spins \cite{Fu2007a,Noh2008a,Hsieh2008,Zhang2009,Hsieh2009,Xia2009,Chen2009,Pan2011c}. 
The number of experimentally identified 3D TIs has grown since the initial discovery, now including the Bi$_{1-x}$Sb$_x$ alloy \cite{Hsieh2008}, Bi$_2$Se$_3$ and Bi$_2$Te$_3$ and their mixed anion variants \cite{Xia2009, Chen2009, xu2010discovery, PhysRevB.82.241306,ji2012bi}, GeBi$_2$Te$_4$ and its PbSe-Bi$_2$Se$_3$ variants \cite{xu2010discovery, nakayama2012manipulation}, TlBiSe$_2$ \cite{sato2010direct, PhysRevLett.105.266401}, and SnTe \cite{nakayama2012manipulation}. Of these materials, Bi$_2$Se$_3$ has been highlighted as a model system due to its large bulk gap and simple surface electronic structure - a single Dirac-like topological surface state (TSS) whose Dirac point resides inside the gap, essentially fully spin polarized and robust against the elastic scattering and scattering on phonons \cite{Zhang2009,Pan2011c,Jozwiak2011,Bianchi2011,Benia2011,Valla2012a,Pan2012}. 

Bi$_2$Se$_3$ is actually only one member of a family of compounds, called an \lq\lq{}infinitely adaptive series\rq\rq{}, between pure Bi metal and Bi$_2$Se$_3$, made by stacking integer numbers of bismuth bi-layers and Bi$_2$Se$_3$ layers, in compounds of the type (Bi$_2$)$_m$(Bi$_2$Se$_3$)$_n$ \cite{lind2003general}. Given the topological properties of the surface of Bi$_2$Se$_3$, and the fact that the bismuth bi-layer has been predicted to be a two-dimensional quantum spin Hall system \cite{Murakami2006}, it is of interest to determine whether the superlattice material also has topological surface states. Thus far, however, these materials have been studied only structurally as small crystals, or in polycrystalline form \cite{lind2003general}. Here we report the successful growth of large single crystals of the simplest member of the infinitely adaptive series, i.e. Bi$_4$Se$_3$ (m=n=1), stabilized for crystal growth by partial S substitution for Se, and present spin- and angle-resolved photoemission spectroscopy (SARPES) studies of its surface electronic structure. Our measurements reveal a hexagonally shaped surface state whose spin structure indicates that Bi$_4$Se$_3$ is a semimetal that supports topological surface states. The compound is the natural bulk analog of recently described artificial material made by Bi adsorption on the surface of Bi$_2$Te$_3$ \cite{Hirahara2011}. Photoemission electron microscopy (PEEM) reveals that the material cleaves with mixed Bi-layer and Se-layer terminations of the exposed terraces. This prohibits the association of the TSS with either the Bi or Se terraces by (S)ARPES, but it suggests that this material offers the unique opportunity for studying novel 1D topological physics near the terrace steps by nanometer-scale probes. 

\begin{figure}[htb]
\begin{center}
\includegraphics[width=8.5cm]{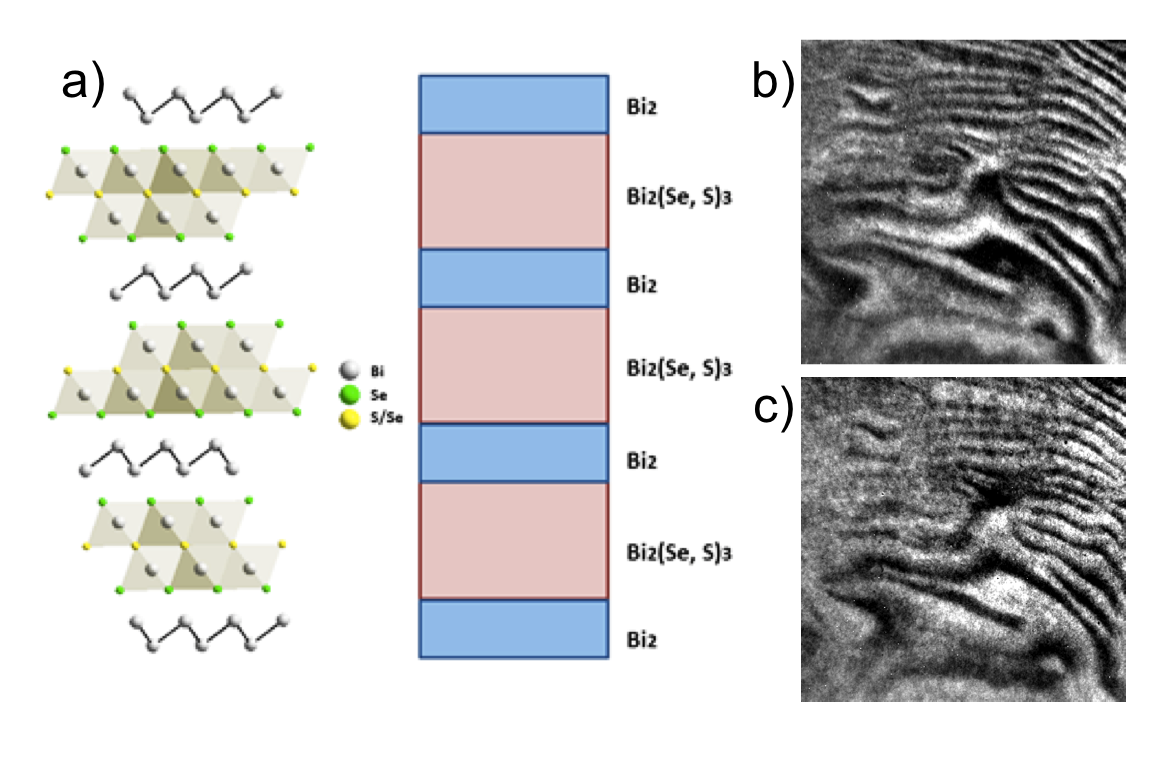}
\caption{(Color on line) Crystal structure (a) and PEEM images of a Bi$_4$Se$_{2.6}$S$_{0.4}$ surface, taken at the Bi 5$d$ (b) and Se 3$d$ (c) core levels with 100 eV photons. Field of view in the PEEM images is 20$\times$20 $\mu$m. Bright (dark) contours reflect regions of high (low) photoemission intensities. 
}
\label{Fig1}
\end{center}
\end{figure}

High purity elemental Bi (99.999\%), Se (99.999\%), and S (99.999\%) were used for the Bi$_4$Se$_{2.6}$S$_{0.4}$ crystal growth, with a starting formula of Bi$_{4.5}$Se$_2$S; the excess bismuth was used as a flux and the S greatly enhances the selective crystal growth of the 4:3 phase. The pre-cleaned elements were sealed in clean quartz ampoules, heated to $850~^{\circ}$C for 1 day, and cooling over the period of 1 hour to $700~^{\circ}$C. The crystal growth involved cooling from $700~^{\circ}$C to $550~^{\circ}$C over a period of 25 h, followed by centrifuging. Single crystals up to 1 cm$^2$ area in the basal plane and several millimeters thick were obtained, with high-luster cleavage faces. Single crystal X-ray diffraction data was collected on a Bruker APEX II diffractometer using Mo K$\alpha$ radiation at room temperature. Unit cell determination and refinement, and data integration were performed with Bruker APEX2 software. The crystal structure was determined using SHELXL-97 implemented through WinGX.

The ARPES experiments were carried out on a Scienta SES-100 electron spectrometer at beamline 12.0.1 of the Advanced Light Source. The spectra were recorded at photon energies ranging from 35 to 100 eV, as indicated in the text, with a combined instrumental energy resolution of $\sim12$ meV and an angular resolution better than $\pm 0.07^{\circ}$. Samples were cleaved at low temperature (15-20 K) under ultra-high vacuum (UHV) conditions. The temperature was measured using a silicon sensor mounted near the sample.
The SARPES experiments were performed at beamline U5UA at the National Synchrotron Light Source using a Scienta R4000 electron analyzer coupled to a pair of mini-Mott spin polarimeters that enable measurements of both the in-plane and out-of plane spin polarizations. The spin-resolved data were recorded at 58 eV photon energy, with the sample kept at $\sim300$ K. Energy and angle resolution were approximately 40 meV and $0.5^{\circ}$, respectively. 
Photoemission Electron Microscopy (PEEM) experiments were carried out at the XPEEM/LEEM end-station at the National Synchrotron Light Source beamline U5UA at room temperature, with 100 eV photons using an Elmitec SPE-LEEM III microscope.
All the samples were cut from the same bulk piece and cleaved and measured in ultrahigh vacuum conditions (base pressure better than $2\times 10^{-9}$ Pa in the ARPES and SARPES chambers and better than $2\times 10^{-8}$ Pa in the PEEM chamber).

Electronic structure calculations where performed in the framework of density functional theory (DFT) using the \textsc{wien2k} \cite{blaha2001} code with a full-potential linearized augmented plane-wave and local orbitals basis \cite{singh2006,madsen2001,sjaestedt_alternative_2000} together with the Perdew-Burke-Ernzerhof parameterization \cite{perdew_generalized_1996} of the generalized gradient approximation for the
exchange-correlation functional. The plane wave cut-off parameter
R$_{MT}$K$_{MAX}$ was set to 7 and the irreducible Brillouin zone (BZ) was sampled by 288 k-points. Spin orbit coupling (SOC) was included. For Bi$_4$Se$_3$ the reported crystal structure \cite{Bi4Se3} was used. To estimate the effect of the partial S substitution for Se, electronic structure calculations were also performed for Bi$_4$Se$_2$S. For this material a fully ordered structure was taken (with S fully occupying the middle layer in the quintuple layer part of the structure) and the lattice constants were interpolated linearly, according to Vegards's law \cite{vegard1921konstitution}, using our data for the structure of Bi$_4$Se$_{2.6}$S$_{0.4}$.

\begin{table}[t]
\begin{center}
\caption{Structural Characterization of Bi$_{4}$Se$_{2.56}$S$_{0.44}$.  Space group R$\bar 3$(No. 148), a = 4.2466(8) $\AA$, c = 39.706(7) $\AA$, 329 unique reflections, R$_1$ (all reflections) = 0.0457, wR$_2$ = 0.0938, GooF = 1.109.}
\tiny
\begin{tabular}{c|c|c|c|c|c|c|c|c}
\hline
\hline
Atom & Wyck. & x & y & z & Occ. & U$_{11}$ & U$_{33}$ & U$_{12}$\\
Bi (1) & 6c & 0 & 0 & 0.14390(3) &  & 0.0227(4)& 0.0307(6) & 0.0114(2)\\
Bi (2) & 6c & 2/3 & 1/3 & 0.04505(2) &  & 0.0200(4)& 0.0262(6) & 0.0100(2)\\
Se (1) & 6c & 1/3 & 2/3 & 0.08384(8) &  & 0.0276(9)& 0.0340(17) & 0.0138(4)\\
Se (2) & 3a & 0 & 0 & 0 & 0.56(2) & 0.024(2)& 0.024(3) & 0.0122(10)\\
S (2) & 3a & 0 & 0 & 0 & 0.44(2) & 0.024(2)& 0.024(3) & 0.0122(10)\\
\hline
\hline
\end{tabular}
\end{center}
\end{table}

Figure \ref{Fig1} shows a schematic view of the crystal structure of Bi$_4$Se$_{2.6}$S$_{0.4}$ and PEEM images of a cleaved surface. The crystal structure of the material was determined quantitatively by least squares structure analysis of a single crystal. The details of the structural characterization are given in Table 1; the structure determination was used to precisely define both the crystal structure and the composition. The two PEEM micrographs show the same spot on the surface recorded at two different kinetic energies of photo-electrons, corresponding to the higher 5/2 component of the Bi 5$d$ (see Fig. 2(a)) and Se 3$d$ core levels. The inversion of contrast indicates that the surface consists of alternating Bi$_2$ and Se terminated terraces. The terrace width is of the order of one $\mu$m. We also note that the lower 5/2 Bi 5$d$ component and the valence band (at $\sim0.5$ eV below the Fermi level) PEEM micrographs have the same contrast as the Se 3$d$ core level, indicating that these features are related to the Se-terminated terraces.
As the photon spot sizes in our ARPES and SARPES experiments are around 50 $\mu$m and 0.5 mm, respectively, the photoelectron characterization represents an average over the two types of terraces.
\begin{figure}[htbp]
\begin{center}
\includegraphics[width=9cm]{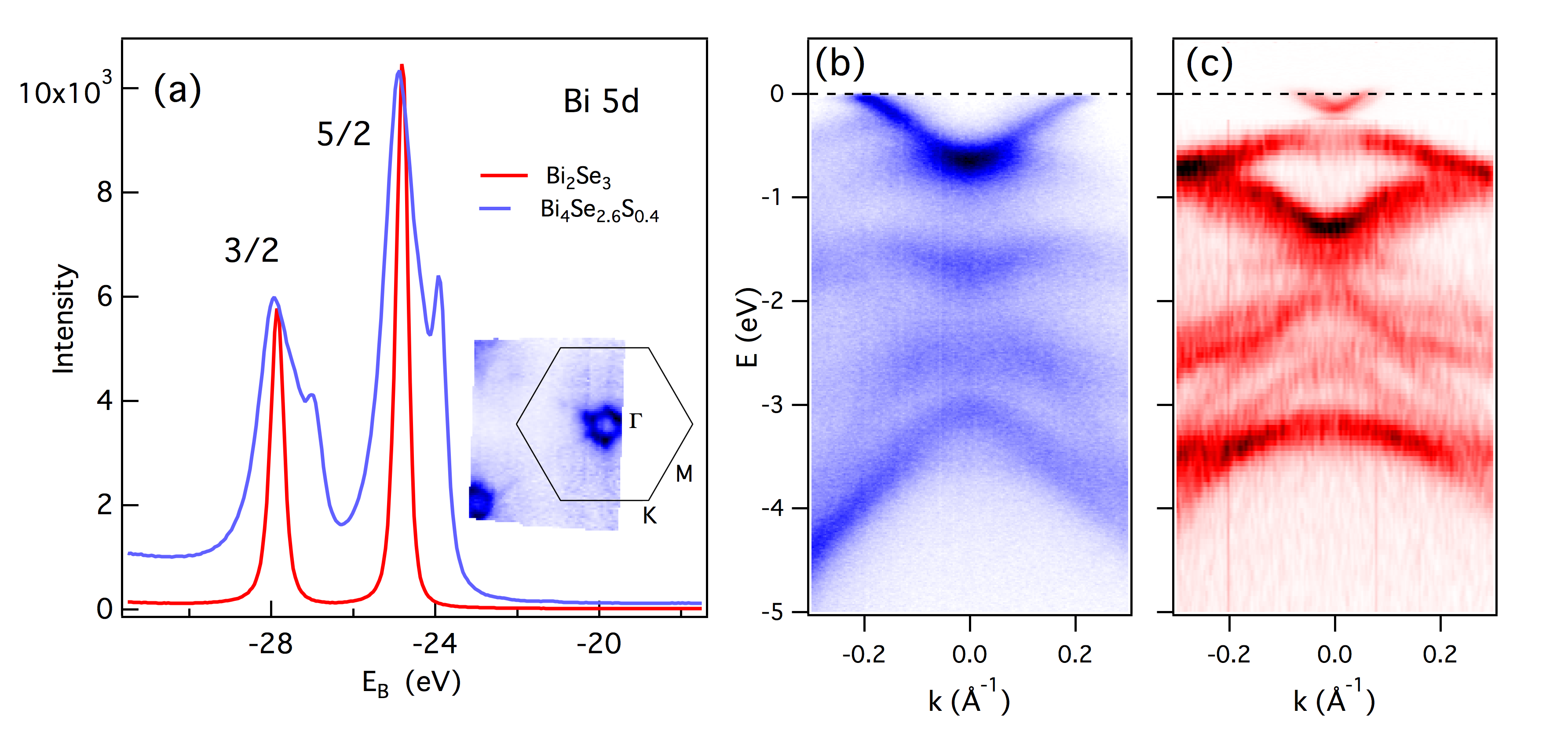}
\caption{(Color on line) Photoemission Spectra from Bi$_4$Se$_{2.6}$S$_{0.4}$ and Bi$_2$Se$_3$. (a) Bi 5$d$ core-levels in Bi$_4$Se$_{2.6}$S$_{0.4}$ (blue) and Bi$_2$Se$_3$ (red). Inset shows the SBZ with the Bi$_4$Se$_{2.6}$S$_{0.4}$ Fermi surface measured with 59 eV photons. (b) Valence band along the $\Gamma-M$ line in the SBZ for Bi$_4$Se$_{2.6}$S$_{0.4}$ and (c) for Bi$_2$Se$_3$.
}
\label{Fig2}
\end{center}
\end{figure}

\begin{figure*}[htb]
\begin{center}
\includegraphics[width=13cm]{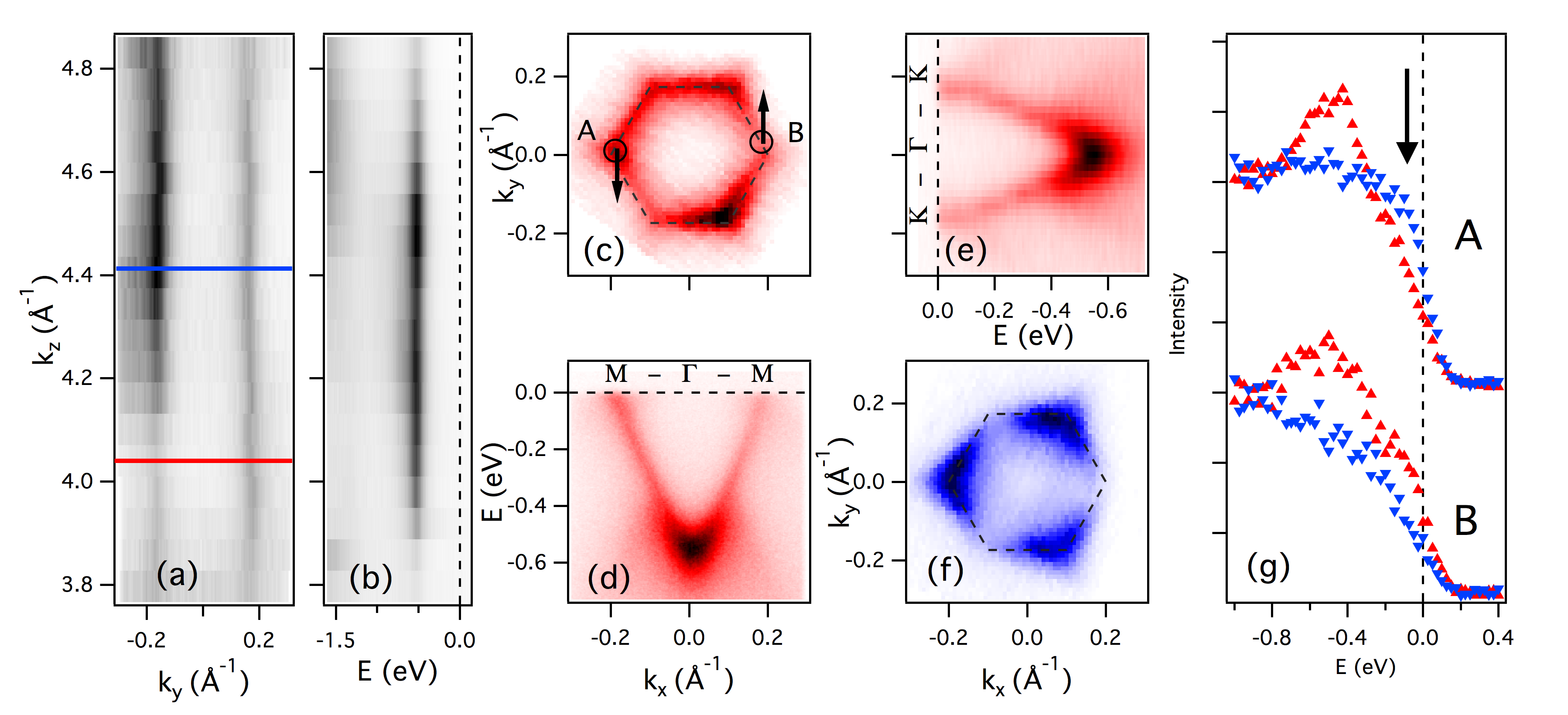}
\caption{ (Color on line) Spin-resolved electronic structure of Bi$_4$Se$_{2.6}$S$_{0.4}$. (a) ARPES intensity at the Fermi level as a function of the in-plane ($k_y$ along the $\Gamma-K$ line) and out-of plane ($k_z$) momentum. (b) Intensity at the $\Gamma$ point ($k_x=k_y=0$) of the SBZ as a function of $k_z$. (c) In-plane Fermi surface measured at $k_z=4.06$ $\mathrm\AA^{-1}$ and (f) at $k_z=4.43$ $\mathrm\AA^{-1}$, corresponding to the red and blue lines in panel (a), respectively. The circles A and B in (c) indicate the $k$-positions where the spin-resolved measurements were performed, while arrows represent the in-plane spin polarization. Dispersion of the surface state along the $\Gamma-M$ (d) and $\Gamma-K$ (e) momentum lines in the SBZ, measured at $k_z=4.06$ $\mathrm\AA^{-1}$. (g) SARPES spectra taken at A and B points in the SBZ, as indicated in (c). Red (blue) spectra represent spin-up (spin-down) components in the $y$ direction. TSS is indicated by arrow.
}
\label{Fig3}
\end{center}
\end{figure*}

In Figure \ref{Fig2} we compare the experimentally determined electronic structures of Bi$_4$Se$_{2.6}$S$_{0.4}$ and Bi$_2$Se$_3$. 
The Bismuth core level (5$d$) spectra (Fig. \ref{Fig2}(a)) show a double peak structure of each spin-orbit split state, clearly indicating two different chemical environments for Bi atoms in Bi$_4$Se$_{2.6}$S$_{0.4}$: the lower one is similar to the environment in Bi$_2$Se$_3$, whereas the higher one indicates the more metallic character of the Bi$_2$ layers. This assignment is in agreement with the PEEM data as these two components display a reversed contrast. The wide-energy range valence band spectra, shown in Figure \ref{Fig2} (b) and (c) are generally similar for the two materials, with a nearly rigid-band down-shift of $\sim 0.4$ eV of the Bi$_4$Se$_{2.6}$S$_{0.4}$ spectrum relative to Bi$_2$Se$_3$ yielding a good general correspondence of the bands. The high intensity feature in the Bi$_4$Se$_{2.6}$S$_{0.4}$ spectrum that crosses the Fermi level is the central point of this study - this is the TSS, similar to the one observed in Bi$_2$Se$_3$, but also down-shifted, with the Dirac point around 0.54 eV below the Fermi level. This down-shift of both the valence bands and the TSS suggests that the Bi$_2$ layers in this natural superlattice phase donate electrons to the Bi$_2$Se$_3$ quintuple layers. A PEEM micrograph taken at $\sim 0.5 eV$ below the Fermi level shows the same contrast as the Se 3$d$ core level, suggesting that the TSS measured in ARPES is residing on Se-terminated terraces. A wide $k$-space survey indicates that the TSS forms a hexagonally shaped Fermi surface at the center of the surface Brillouin zone (SBZ) (Fig. \ref{Fig2}(a), inset). A new feature in the valence band spectrum in Fig. \ref{Fig2}(b) is the \lq\lq{}hole-like\rq\rq{} state that crosses with the TSS at E$\approx -0.15$ eV, $k_x\approx\pm 0.15$\AA$^{-1}$. This is a bulk state, as will be discussed further with reference to Fig. \ref{Fig3}, which does not have an obvious counterpart in the Bi$_2$Se$_3$ spectrum.

In Figure \ref{Fig3}, we show the detailed spin- and $k$- resolved electronic structure of Bi$_4$Se$_{2.6}$S$_{0.4}$. Panels (a) and (b) show the photon energy dependence of the ARPES intensity at the Fermi level along the $\Gamma-K$ line and at the center of the SBZ, respectively, for photon energies ranging from 50 to 84 eV in 2 eV steps. The intensity maps are converted into $k$-space by using the inner potential $V = 10$ eV in the approximation for the out-of-plane momentum, $k_{z}=\frac{1}{\hbar}\sqrt{2m(E_{k}cos^{2}(\theta)+V)}$, where $E_k$ is the kinetic energy of a photoelectron. The straight features at $k_y = \pm 0.173$ \AA$^{-1}$ in Fig. \ref{Fig3}(a) represent the Fermi surface of the TSS. The bulk hole state previously mentioned is visible here for negative $k_y$ as a slightly dispersing (with $k_z$) feature, intertwined around the TSS. The straight vertical line of high intensity in (b) corresponds to the Dirac point of the TSS. Slightly lower in energy, another bulk state can be seen as dispersing with $k_z$, with the minimum around $k_z=4.35$ \AA$^{-1}$. Both bulk features have the periodicity of $\approx0.5$ \AA$^{-1}$, suggesting that the relevant periodicity in real space in the $z$ direction is $\approx12.5$ \AA. This is close to the thickness of one of the Bi$_2$-Bi$_2$Se$_3$ composite layers that repeats three times to form the rhombohedral unit cell (Fig. \ref{Fig1}(a)). In panels (c) and (f), we also show the in-plane ARPES intensity at the Fermi level for two photon energies, $h\nu=58$ and 72 eV, corresponding to $k_z=4.06$ and 4.43 $\mathrm\AA^{-1}$, respectively. 
The hexagonally shaped contour represents the Fermi surface of the TSS. The dispersion of the surface state measured at $h\nu=58$ eV along two high symmetry directions in the SBZ is shown in panels (d) and (e). At $h\nu=72$ eV, the intensity at the Fermi level is strongly modulated by the nearly overlapping bulk state with the three-fold symmetry (Fig. \ref{Fig3}(f)). The area enclosed by the TSS\rq{} Fermi surface is almost an order of magnitude larger than that in pristine Bi$_2$Se$_3$, reflecting a surface electron concentration of $\sim8.7\times10^{12}$ cm$^{-2}$. Such high surface electron doping can also be obtained on Bi$_2$Se$_3$ through adsorption of various metals on it\rq{}s surface, producing a very similar hexagonal warping of the TSS\rq{} Fermi surface \cite{Valla2012a}. 

Figure \ref{Fig3}(g) shows the spin-resolved spectra with polarization defined in the $y$ direction for points A and B from Fig. \ref{Fig3}(c). 
The peak at the Fermi level, corresponding to the TSS, shows a significant spin polarization ($\sim30\%$) that reverses sign when the momentum is reversed ($k_x \rightarrow -k_x$), an indication of the chiral spin structure. It is interesting to note that the chirality is opposite to that observed in Bi$_2$Se$_3$ \cite{Pan2011c}. The peak at around $-0.5$ eV also shows a polarization, but without the sign-reversal, an effect also observed in the bulk valence band in Bi$_2$Se$_3$ \cite{Jozwiak2011}. The polarization of the TSS is much smaller in Bi$_4$Se$_{2.6}$S$_{0.4}$ than in Bi$_2$Se$_3$ \cite{Pan2011c}, probably due to the partial overlap with the bulk states or with the Bi$_2$-derived states.  

\begin{figure}[t]
\begin{center}
\includegraphics[width=7cm]{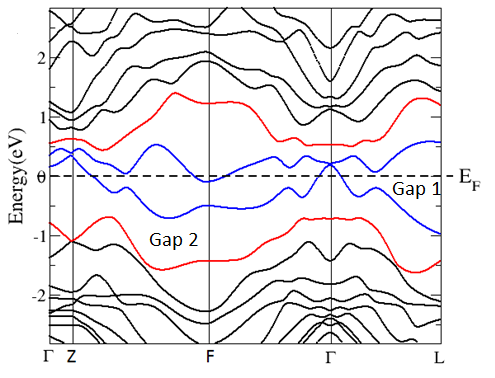}
\caption{(Color on line) Bulk electronic band structure of Bi$_4$Se$_3$. Colored bands are described in the text. 
}
\label{Fig4}
\end{center}
\end{figure}

Figure \ref{Fig4} shows the calculated bulk band structure of Bi$_4$Se$_3$. SOC plays a significant role as, for example, the lower red colored band has shifted by about 1~eV on the inclusion of SOC. The calculations predict Bi$_4$Se$_3$ to be a semi metal with hole pockets at $\Gamma$ and Z and an electron pocket at F. The hole pockets at $\Gamma$ and Z project to the $\bar{\Gamma}$ point in the SBZ and are seen in the ARPES measurement. The electron pocket predicted at the bulk F point is not observed by ARPES, suggesting that it lies higher in energy than is predicted by the calculations. Comparison with the Bi$_2$Se$_3$ band structure allows us to generally identify bands closely above and below E$_F$ in Bi$_4$Se$_3$ as being Bi$_2$Se$_3$-like bands that have been modified through interactions with the Bi bilayers; these are shown as red in the figure. Bands appearing directly around E$_F$ (blue) in Bi$_4$Se$_3$ are primarily Bi$_2$ derived states modified by hybridization with the Bi$_2$Se$_3$ layers.   

There are two continuous gaps in the electronic structure for Bi$_4$Se$_3$, labeled as Gap 1 and Gap 2 in Figure \ref{Fig4}. Counting all bands to Gap 1 gives, for the time-reversal-invariant momenta (TRIM) points of the bulk BZ \cite{fu2007topological}, positive parity for $\Gamma$, Z, and L, and negative parity for F. Therefore in Bi$_4$Se$_3$ we expect a TSS cone at the projection of the F point in the SBZ. According to the band structure calculation, the cone would be about 0.5~eV below the Fermi energy (for Bi$_4$Se$_2$S we expect a second cone at the projection of the Z point, which also has negative parity, but since this would be above the Fermi level it would not be seen by ARPES). Counting the bands until Gap 2 yields, for both Bi$_4$Se$_3$ and Bi$_4$Se$_2$S, TRIM points Z, M, and L with positive party and $\Gamma$ with negative parity. This would yield a TSS cone at 0.5~eV below the Fermi level at $\Gamma $. Note that counting to Gap 2 is equivalent to counting to the lower red band belonging to Bi$_2$Se$_3$; we observe the same parity at each TRIM as is seen in pure Bi$_2$Se$_3$. When counting to Gap 1 one counts to a band belonging to the Bi bi-layer. The second cone at F might belong to the Bi$_2$, which was predicted to be a 2D quantum spin Hall system \cite{Murakami2006}. The crystal cleaves to expose Bi$_2$ and Bi$_2$Se$_3$ Se-terminated terraces, neither of which are at a bulk inversion center. In this case the projection of the bulk BZ points to the surface BZ is less certain, but since the only TRIM at the surface BZ are at $\bar{\Gamma}$ and $\bar{M}$, the TSS must be at either of those two points. In ARPES we see one TSS cone at $\bar{\Gamma}$ and no cone at $\bar{M}$ (Fig. \ref{Fig2}(a), inset). Where these surface states reside, i.e. on the Bi$_2$ terraces, the Bi$_2$Se$_3$ terraces, or both, and their relation to the bulk band structure, should be the subject of further study on this material.     

Bi$_4$Se$_{2.6}$S$_{0.4}$ opens several interesting possibilities for studying the role of dimensionality in topological systems: as Fig. \ref{Fig3} shows, there is a 2D n-type TSS, nearly overlapping with a p-type bulk (3D) Fermi surface. In addition, PEEM images from Fig.\ref{Fig1} show alternating Bi$_2$ - Se terminations, each kind of surface no doubt displaying a different surface electronic structure.  Each step would then represent a boundary (or a junction) between two 2D (topological-\rq{}trivial\rq{} or topological-topological, n-n or p-n) electronic \lq\lq{}gasses\rq\rq{}. This offers the possibility of studying topological phenomena ranging from 1D to 2D to 3D in a single system. The current ARPES experiments lack the necessary spatial resolution to observe such effects, but the scanning tunneling microscopy, for example, could provide information about these phenomena.
A recent study of deposited Bi bi-layer on Bi$_2$Te$_3$ \cite{Hirahara2011} has indicated that similar phenomena might be at play in that system: with depositing bismuth, a hole-like state is added to the Bi$_2$Te$_3$ bands. Although the authors consider the Bi$_2$ layer to be homogeneous, it might be that terraces with mixed termination are formed in that case as well, creating a situation that is similar to what we observe in Bi$_4$Se$_{2.6}$S$_{0.4}$.

We acknowledge helpful discussions with B.A. Bernevig and C. Fang.   
The work at Brookhaven is supported by the US Department of Energy (DOE) under Contract No. DE-AC02-98CH10886. The work at Princeton University was supported by SPAWAR grant N66001-11-1-4110 and NSF DMR-0819860, and HWJ thanks ConocoPhillips for fellowship support.  
ALS is operated by the US DOE under Contract No. DE-AC03-76SF00098.

\end{document}